\documentclass[runningheads]{svmult}

\usepackage{makeidx}   
\usepackage{graphicx}  
\usepackage{subeqnar}  
\usepackage{multicol}  
\usepackage{physprbb}  
\makeindex             

\begin{document}
\title*{Activity and Kinematics of M and L Dwarfs\protect\newline}
\toctitle{Activity and Kinematics of M and L Dwarfs
\protect\newline in the Particle Deflection Plane}
%
%
\titlerunning{Activity and Kinematics}
%
\author{John E. Gizis}

\authorrunning{John E. Gizis}
%
%
\institute{Infrared Processing and Analysis Center\\
California Institute of Technology\\
Pasadena CA 91125 USA}

\maketitle              

\begin{abstract}
I discuss observations of two traditional age indicators,
chromospheric activity and kinematics,
in late-M and L dwarfs near the hydrogen-burning limit.
The frequency and strength of chromospheric activity disappears 
rapidly as a function of temperature over spectral types M8-L4.
There is evidence that young late-M and L dwarfs have weaker
activity than older ones, the opposite of the traditional
stellar age-activity relation.
The kinematics of L dwarfs confirm that lithium L dwarfs are
younger than non-lithium dwarfs.   
\end{abstract}

\section{Introduction}

The Two Micron All-Sky Survey (2MASS) has enabled large
samples of cool dwarfs to be detected and studied.  
Over the spectral range M8 to L8, the
traditional TiO and VO molecular bands disappear, grains
become important, and the appearance of the spectrum changes drastically
\cite{k99}.  

\begin{figure}[htbp]
\begin{center}
\rotatebox{270}{
\includegraphics[width=.6\textwidth]{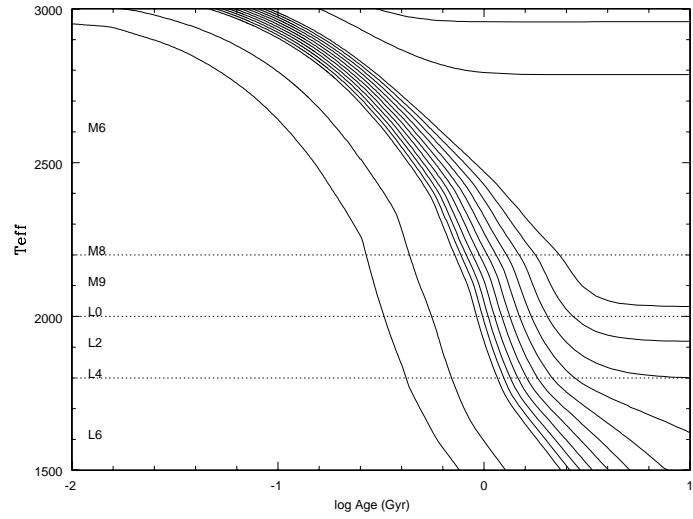}
}
\end{center}
\caption[]{Theoretical model tracks for stars and brown dwarfs
\cite{burrows97}:  Age
information allows stars and brown dwarfs to be distinguished.}
\label{fig_teffage}
\end{figure}

\begin{figure}[hptb]
\begin{center}
\rotatebox{270}{
\includegraphics[width=.6\textwidth]{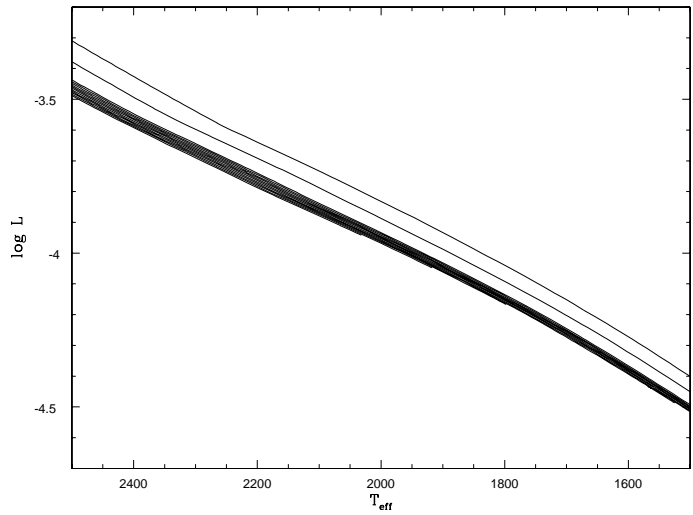}
}
\end{center}
\caption[]{Theoretical model tracks for stars and brown dwarfs 
\cite{burrows97}:  
Temperatures and luminosities do not provide mass or age information.}
\label{fig_tefflum}
\end{figure}

This spectral range, corresponding to temperatures
between $\sim 2300$ and $\sim 1400$K, covers the crucial
transition between hydrogen-burning stars and 
brown dwarfs.  The situation is illustrated in
Figs.~\ref{fig_teffage} and~\ref{fig_tefflum} using 
theoretical models \cite{burrows97} and
a plausible temperature scale \cite{k99}.  
(See this volume for different views of the L dwarf temperature scale).
Both stars and brown dwarfs cool as they contract.   Stars, however,
eventually stabilize at a constant temperature (the upper
tracks in Fig.~\ref{fig_teffage}), while brown dwarfs continue to cool
(the lower tracks).  Current theoretical models
suggest that $0.075 M_\odot$ stars may exist as cool as $\sim 1800$K
\cite{burrows97,bcah98}, corresponding to $\sim$L4.  (In the hotter
L dwarf temperature scales, even later L dwarfs may be stars.)

On the basis of Fig.~\ref{fig_teffage}, it is clear that field 
brown dwarfs and stars
could be easily distinguished if we could measure age -- but
age is not directly measurable and must be deduced
from indirect measures.  
Luminosities (in the form of absolute magnitudes) and 
temperatures (in the form of colors and spectral types),
{\it can} easily be measured, but as seen in Fig.~\ref{fig_tefflum}, different
mass objects follow evolutionary tracks that are practically
indistinguishable.

The need for an age indicator for M and L dwarfs motivates an investigation of
two traditional stellar age indicators.  Chromospheric
activity is linked to age in convective stars through the dynamo:
a star is born rapidly rotating, the rotation drives a dynamo
which produces magnetic fields, the fields produce
activity (chromosphere, corona, flares, and wind)
the wind spins down the star, the dynamo is weakened,
and the observable activity then decreases with age.    
Kinematics are also linked to age:  Stars are born
from molecular clouds with low random space velocities,
but with time encounters other stars and clouds in the
Galactic disk 'heat' the stellar velocity distribution.  
Most of the discussion in this paper is drawn from
Gizis et al. \cite{gizis2000}.  

\section{M and L Dwarf Observations}

\begin{figure}[hptb]
\begin{center}
\rotatebox{270}{
\includegraphics[width=.6\textwidth]{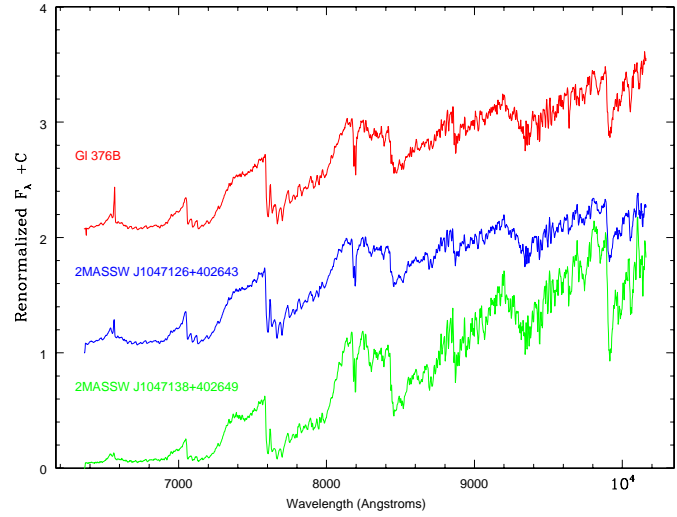}
}
\end{center}
\caption[]{Spectra of late-M dwarfs illustrating the richness
of molecular features and the H$\alpha$ emission line at 6563\AA.
The M8 dwarf 2MASSW J1047138+402629 (LP 213-68) has much 
weaker activity than its M6.5 primary 2MASSW J1047126+402643 (LP 213-67) 
\cite{gizis2000a}.}
\label{fig_mspectra}
\end{figure}

\begin{figure}[hptb]
\begin{center}
\rotatebox{270}{
\includegraphics[width=.6\textwidth]{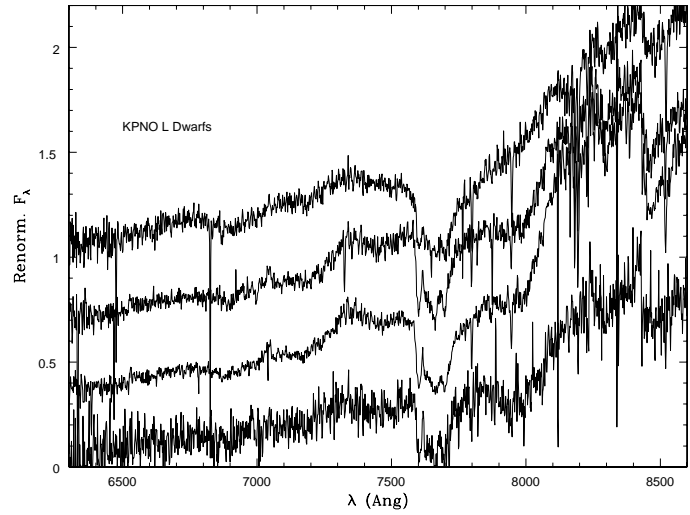}
}
\end{center}
\caption[]{KPNO 4-meter spectra of 4 nearby L dwarfs with 
K$_s<12$.  Even in modest exposure times ($<30$ minutes), these
L dwarfs are accessible to 4-meter telescopes.}
\label{fig_lspectra}
\end{figure}

\begin{figure}[hptb]
\begin{center}
\rotatebox{270}{
\includegraphics[width=.6\textwidth]{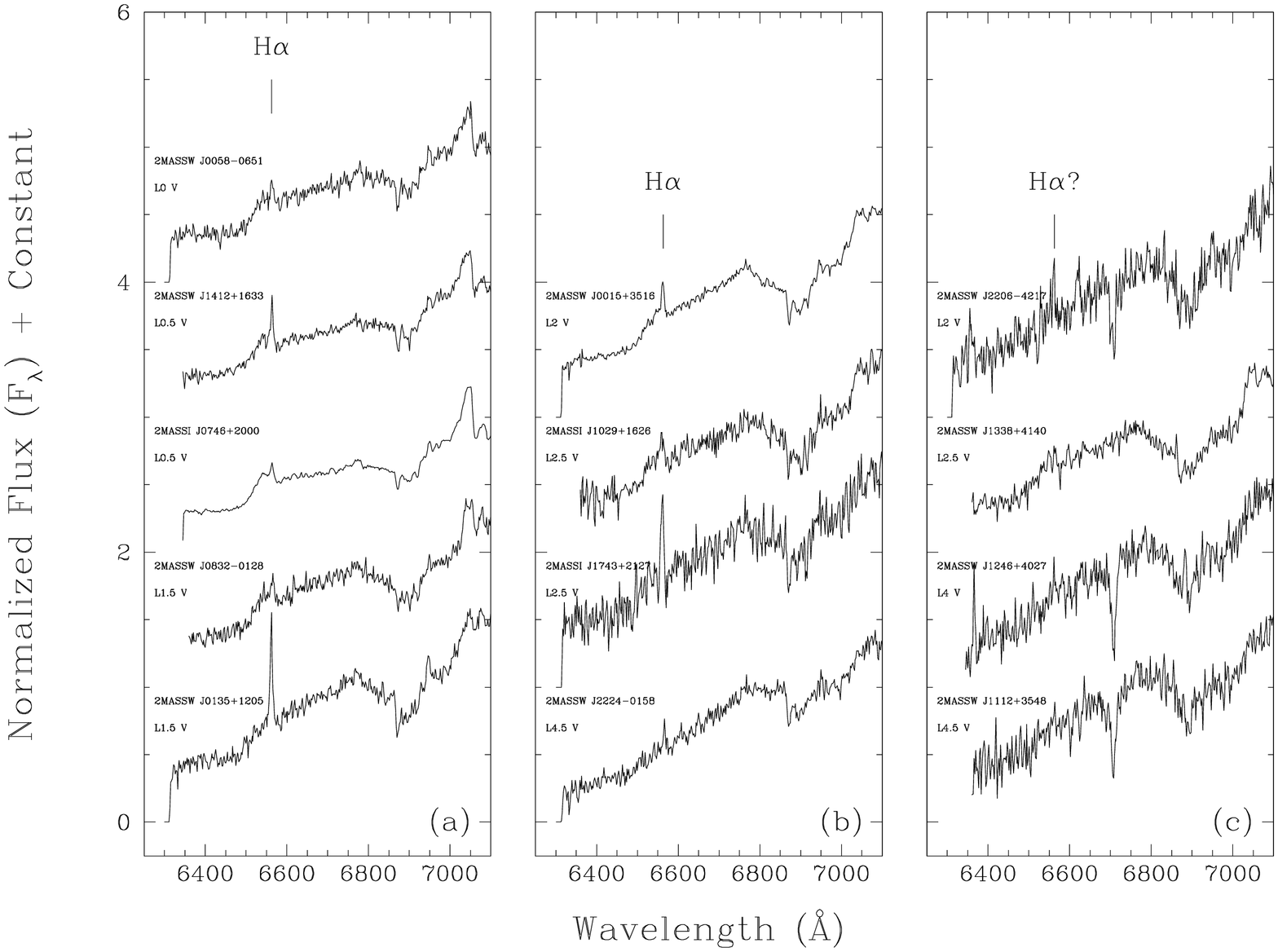}
}
\end{center}
\caption[]{Keck spectra of L dwarfs from Kirkpatrick et al.\cite{k00}.  
Low (9\AA) resolution spectra
allows both the H$\alpha$ (6563\AA) emission line and lithium 
(6708\AA) absorption line
to be observed.  L dwarfs with lithium, although young, generally
lack H$\alpha$ emission.  
}
\label{fig_keckha}
\end{figure}

Two large samples form the basis of the discussion in
this paper.  Late-M and early-L dwarfs have been selected to 
$K_s<12$ on the basis of their colors without
any kinematic bias \cite{gizis2000}.  This sample is
supplemented by the large sample of fainter L dwarfs
also selected on the basis of color \cite{k00}.  Observationally,
a sample of nearby bright M and L dwarfs offers considerable
advantages.  They are bright enough to relatively easily 
obtain far-red spectroscopy ($>6000$\AA), allowing
the measurement of spectral types, H$\alpha$ emission,
lithium absorption, surface gravity, and a photometric
parallax (Figs.~\ref{fig_mspectra},~\ref{fig_lspectra},
and~\ref{fig_keckha}).  
This can be supplemented by astrometry, providing
proper motions, trigonometric
parallaxes, and hence tangential velocities.  In the case
of the late-M dwarfs, the sample is bright enough that
the Palomar plates plus 2MASS allow measurements of the
proper motions, which Gizis et al.\cite{gizis2000} combined with 
the photometric distance estimate to obtain $v_{tan}$. 
The Kirkpatrick et al.\cite{k99,k00} L dwarfs are faint enough that this is
not possible, but United States Naval Observatory has
measured proper motions and trigonometric parallaxes for
a representative sample using CCD astrometry.  Together, near-infrared 
photometry, spectroscopy and astrometry provide a rich
set of diagnostics.  The discovery of additional bright L dwarfs
with $K_s<12$ will aid many studies, such as 
3.3 micron observations of methane (Noll, this volume) and 
companion searches (Reid, this volume).  

\section{Activity}

The 2MASS samples allow an initial reconnaissance of chromospheric
activity in M and L dwarfs.  Two critical questions may be 
addressed:  How {\it frequent} is chromospheric activity 
in cool dwarfs?  How {\it strong} is chromospheric activity?

\begin{figure}[hptb]
\begin{center}
\rotatebox{270}{
\includegraphics[width=.6\textwidth]{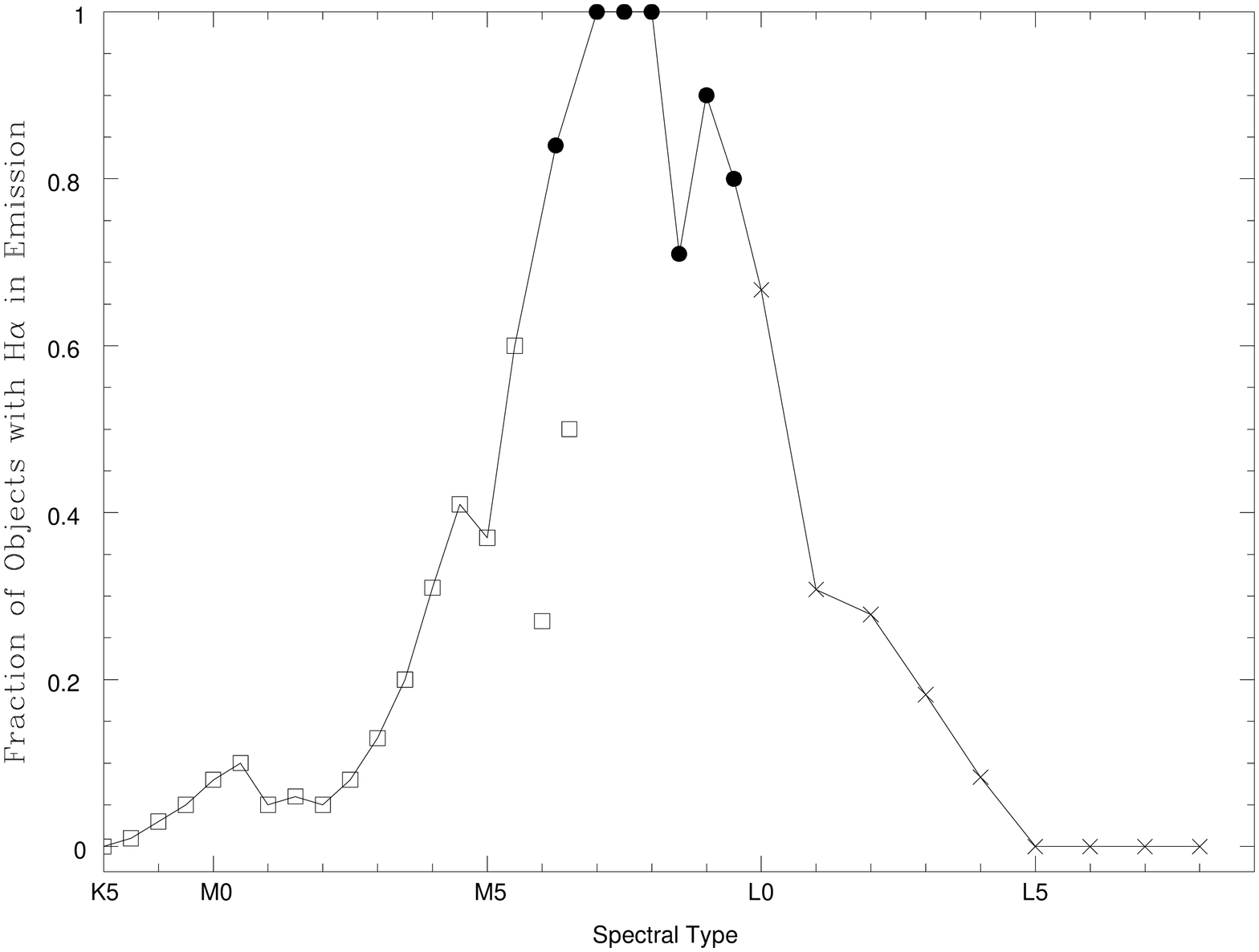}
}
\end{center}
\caption[]{The frequency of H$\alpha$ emission for nearby field 
M and L dwarfs\cite{gizis2000,hgr96,k00}.}
\label{fig_percentha}
\end{figure}

By adding the spectroscopic survey of nearby stars 
by Hawley et al. \cite{hgr96}, the frequency of activity
from K7 ($T_{eff} \approx 4200$K down to L8 ($T_{eff} \approx
1400$K?) is plotted in Fig.~\ref{fig_percentha}.  Activity 
is defined by H$\alpha$ in emission with equivalent width 
$>1$\AA.  Two trends are evident.  Over the (stellar) range 
K7 to M7, the frequency of activity increases as cooler,
lower-mass stars are considered.  The kinematics of 
field M dwarfs \cite{hgr96} and observations of open
clusters \cite{hawley} both indicate that the active stars
are younger than the inactive stars, and that the increase
in the frequency of emission reflects the longer lifetime
of high activity levels.

The new result by Gizis et al.\cite{gizis2000} is to add
the cooler (M7-L8) dwarfs, allowing dwarfs at and
below the hydrogen-burning limit to be examined. 
This reveals (Fig.~\ref{fig_percentha}) that the frequency
of emission declines rapidly from spectral type M8
down to L4 --- with cooler dwarfs in these surveys 
all lacking chromospheric activity.  
As is evident in Fig.~\ref{fig_teffage}, over the range M8 
to L4 an increasing fraction of objects will be young
brown dwarfs rather than old stars.  

\begin{figure}[hptb]
\begin{center}
\rotatebox{270}{
\includegraphics[width=.6\textwidth]{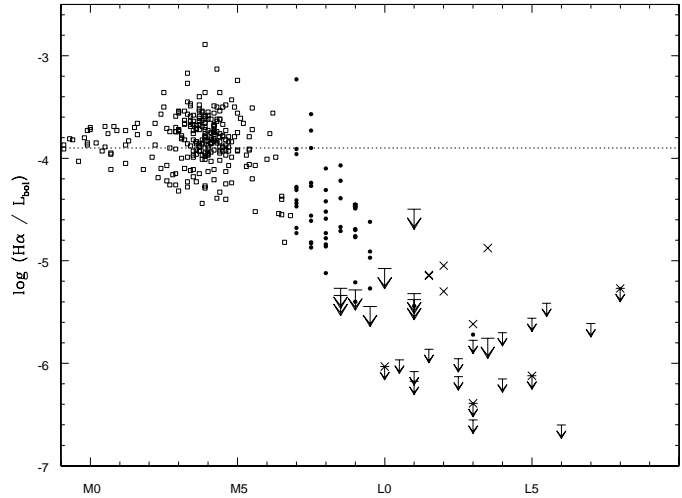}
}
\end{center}
\caption[]{Activity levels as a function of spectral type for M and L dwarfs.
The dotted line 
at -3.9 is the level at which any M dwarf would be observed in emission.
None of the M8 or later dwarfs have activity levels above the -3.9 level,
even though such activity is common in mid-M dwarf stars.  }
\label{fig_sphabol}
\end{figure}

The mere presence or absence of detectable H$\alpha$ emission 
already reveals interesting behavior, but the declining 
photospheric emission near 6500\AA~means that  H$\alpha$
is increasingly easy to detect and equivalent width is 
a poor measure of H$\alpha$'s importance.  We therefore plot the
{\it strength} of H$\alpha$ emission by considering the
ratio of the H$\alpha$ luminosity to the bolometric 
luminosity in Fig.~\ref{fig_sphabol}.  The horizontal
dotted line indicates the level at which 
Hawley et al.\cite{hgr96} could detect activity in
any M dwarf.  It is clear that beyond spectral type M7, 
even the most active dwarfs would not be considered 
very strong by the standards of early-to-mid M dwarfs.
Furthermore, there is a striking temperature dependence, with
the upper envelope of observed activity levels falling
by two orders of magnitude over the range M7 to L4.  This
suggests that effective temperature is the dominant
parameter in controlling the activity levels of these
cool dwarfs.  

\section{Adding Age Information}

Kinematics provide age information for the M and L dwarf
sample.  Unlike activity-age relations, where brown dwarfs
might not obey the same relations as stars, the ``heating''
of the brown dwarf velocity distribution should proceed in the
same way as for stars 
(see \cite{binney} for the theory of this process).  
In particular, young brown dwarfs will tend to have low 
velocities.  The disadvantage of kinematics as an age
indicator is that it is only statistical --- an old
star can have a low random space velocity, with the most notable
example being the Sun itself! 

\begin{figure}[hptb]
\begin{center}
\rotatebox{270}{
\includegraphics[width=.6\textwidth]{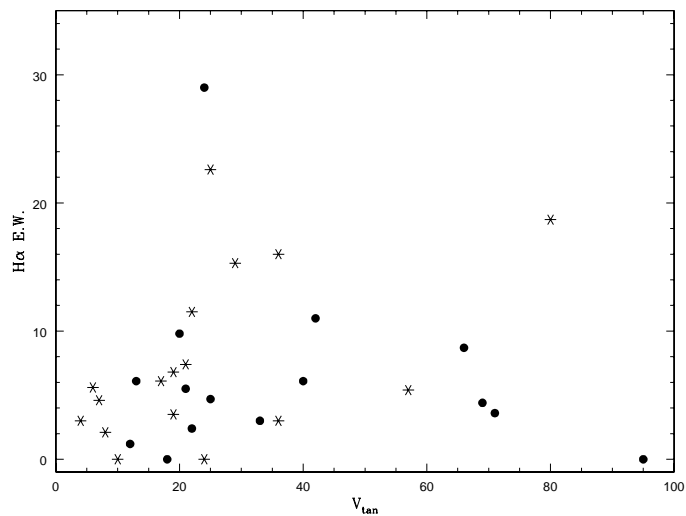}
}
\end{center}
\caption[]{Emission strength and tangential velocities for
M8.0/M8.5 dwarfs (six-pointed stars) and M9.0/M9.5 dwarfs
(solid circles).  The low-velocity, inactive M dwarfs are
a younger population than the high-velocity, active M dwarfs.}
\label{fig_vtanha}
\end{figure}

The tangential velocities of M8 and M9 dwarfs are compared to
their chromospheric activity in Fig.~\ref{fig_vtanha}.  It is
striking that the expected stellar age-activity relation
is not followed.  The population of low velocity ($v_{tan} < 20$ km/s),
hence younger, M8 and M9 dwarfs
have much {\it weaker} emission than the older population
with $v_{tan} > 20$ km/s.  One obvious explanation would be
that the activity levels of the coolest M dwarfs {\it increase}
with age.  However, the theoretical calculations shown in 
Fig.~\ref{fig_teffage} suggest a different explanation.
The young population (corresponding to ages $< \sim 1$ Gyr)
are lower-mass stars and brown dwarfs still in their initial
cooling phase, while the older, more active population must
be hydrogen-burning stars which have largely stabilized
at their main-sequence temperature.  

For L dwarfs, there is another age and mass diagnostic
available in the form of the lithium line\cite{lithium}. 
Objects below $\sim 0.06 M_\odot$ are never hot enough
to destroy lithium and therefore will show a lithium
absorption feature, while more massive stars and
brown dwarfs will burn up their lithium in nuclear reactions
and therefore lack the feature.  At a given temperature
(spectral type), L dwarfs with lithium are thus lower-mass
and younger than L dwarfs without it (see Fig.~\ref{fig_teffage}).
Kirkpatrick et al.'s \cite{k99,k00} data include both
lithium and H$\alpha$ measurements.  
Consider the L1-L4.5 dwarfs, where lithium is detectable even at
low resolution. Only one early-L dwarf, Kelu 1, shows both H$\alpha$ 
emission and lithium
absorption. Eleven other such early-L dwarfs show H$\alpha$ emission but 
do not have lithium absorption. Twelve early-L dwarfs show
lithium absorption but do not have H$\alpha$ emission 
(four of these have marginal H detections or noise consistent with
emission of less than 2 Å). While many L1-L4.5 dwarfs have neither 
H$\alpha$ emission nor lithium absorption, it seems clear that
the chromospherically active L dwarfs are drawn from an older, 
more massive population than the lithium L dwarfs.
Beyond L4.5, there are no definite cases of H$\alpha$ emission, 
although lithium absorption is present for 50\% of the L dwarfs. 
Overall, the L dwarfs show the same properties as 
the coolest M dwarfs --- the younger, less massive 
dwarfs are {\it less} active than the older, more massive
dwarfs, the opposite of the stellar age-activity relation.  

These samples, then, lead to a new view of activity near the 
hydrogen-burning limit.  The dominant parameter is
effective temperature, with cooler dwarfs,
whether stars or brown dwarfs, only being able
to maintain lower activity levels.  The mass is 
a secondary parameter, with lower-mass objects have
weaker activity at a given temperature, at least for
ages typical of field objects.  The importance of 
age, rotation rate, and other parameters needs
further investigation.  

\begin{figure}[hptb]
\begin{center}
\rotatebox{270}{
\includegraphics[width=.6\textwidth]{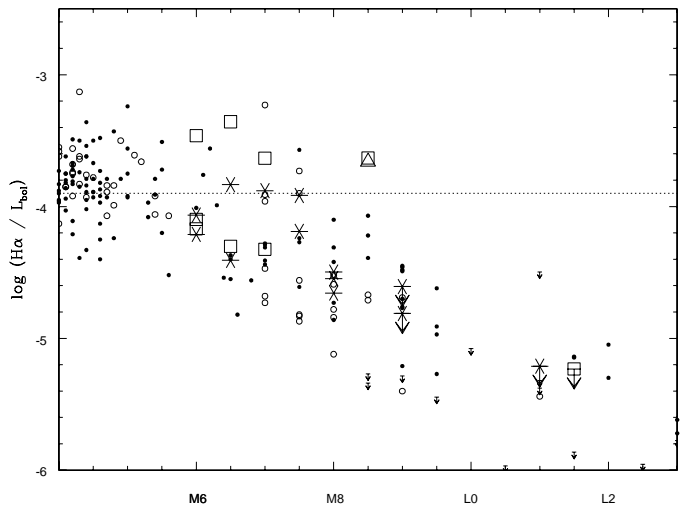}
}
\end{center}
\caption[]{The H$\alpha$ luminosity relative to the bolometric luminosity as a
function of spectral type for both cluster brown dwarfs and field
dwarfs.  Brown dwarfs from the $\sigma$ Ori cluster ($<10^7$ years),
$\rho$ Oph ($<10^7$ years), Pleiades ($\sim 10^8$ years) are shown
as open squares, open triangles, and six-pointed stars respectively.
Note that both cluster L dwarfs have only upper limits on the
detected H$\alpha$ emission.  The field M dwarfs are 
plotted as open circles if $v_{tan}< 20$ km/s and solid
circles for higher velocities.}
\label{fig_sphabol2}
\end{figure}

This new view based on 2MASS field dwarfs can be supplemented
with observations of cluster brown dwarfs of known age.
In Fig.~\ref{fig_sphabol2}, we plot the activity levels 
for Pleiades brown dwarfs \cite{mrzo96,pl2,pl3,zopl} of age $\sim 100$ Myr
with six-pointed stars.  Extremely young brown dwarfs
from the $\rho$Oph \cite{luhman} and the
$\sigma$Ori \cite{sigori1,sigori2} are also shown.  
For comparison, the (old) field dwarfs with
$v_{tan} > 20$ km/s are plotted as solid circles and 
the (young) field dwarfs with $v_{tan} < 20$ km/s
are plotted as open circles.

Despite their youth, activity in the Pleiades brown dwarfs obey
the same fall-off with temperature seen in the field dwarfs.
Their youth is evidently offset by their lower mass, resulting
in unremarkable activity levels.  Indeed, no emission is
detectable in the Pleiades L dwarf, even though activity
is seen in some field L dwarfs of similar spectral type.  

The very young ($<10$ Myr) brown dwarfs show that the relation
between H$\alpha$ emission and age is not simple.  Half of the
young cluster brown dwarfs show weak emission, consistent with the
field M and L dwarfs, but half show much stronger emission.
Like the Pleiades L brown dwarf, the $\sigma$ Ori L-type brown 
dwarf has no detectable activity despite its youth.  In the
case of the $\rho$Oph brown dwarf, the mid-IR excess strongly
suggests that the emission is due to a disk\cite{luhman}, and
Gizis et al.\cite{gizis2000} suggest that the 'active' brown dwarfs
are actually those with disks, while the relatively inactive
brown dwarfs simply show the weak chromospheres expected by analogy with 
the field dwarfs.  

\begin{figure}[hptb]
\begin{center}
\rotatebox{270}{
\includegraphics[width=.6\textwidth]{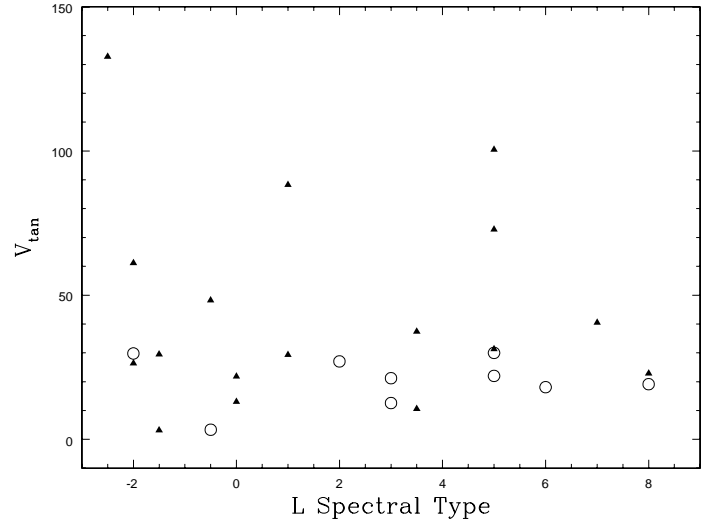}
}
\end{center}
\caption[]{Tangential velocities from U.S.N.O. astrometry
for L dwarfs with (open circles) and without (solid triangles)
lithium.  The lithium L dwarfs have smaller velocities and
hence are a younger propulation.}
\label{fig_ldwarfvtan}
\end{figure}

As kinematic information becomes available for more brown dwarfs,
it will be possible to test evolutionary models statistically.
A first step is shown in Fig.~\ref{fig_ldwarfvtan}, where
tangential velocity is plotted as a function of spectral type.
The tangential velocities are based on USNO parallaxes and
proper motions (Dahn, priv. comm.).  Objects with lithium
are shown as open circles, those without lithium are shown as
solid triangles.  The results are reassuring --- the lithium
L dwarf population is clearly kinematically cooler, hence younger,
than the non-lithium L dwarfs, confirming theoretical expectations.
A larger sample will enable trends with spectral type, activity,
or other properties to be investigated.  Addition of radial
velocities will allow the full three-dimensional space velocity
distribution to be considered.

\section{Further Considerations on Magnetic Activity}

Theoretical investigations are needed to understand the
temperature and mass dependence of H$\alpha$ emission
from the chromospheres of L dwarfs.  Does the
lack of emission mean that there is no chromosphere?  The latest
X-ray observations of an M8 star \cite{vb10} and M9 brown dwarf
\cite{lp944-20} indicate that quiscent X-ray coronae
are not detectable.  The disappearance of coronae is consistent
with the observed weakening of the chromosphere.  Clearly
X-ray observations of L dwarfs would be of great interest
to confirm this trend.

Although the quiscent activity is already disappearing
at M8-M9 spectral types, it is clear that magnetic
fields have not entirely disappeared.  In Gizis et al.'s
\cite{gizis2000} spectroscopic survey, 7\% of the late-M
dwarfs were caught in a flare event.  Liebert et al.\cite{superflare}
have reported a huge flare in a 2MASS M9.5 dwarf.
Reid et al.\cite{ourbri0021} observed a flare in the
well-known M9.5 dwarf BRI0021, the prototype 'inactive'
rapid-rotator \cite{bm95}, and also estimate a 7\% flare rate.   
Flaring activity apparently occurs in objects both with
and without quiscent chromospheric emission.  X-ray flares 
have also been observed in the M9 brown dwarf LP944-20
\cite{lp944-20} with a similar estimated flare rate.  
Despite the weakening of both the chromosphere and corona,
flaring is frequent and strong in the coolest M dwarfs, suggesting
that magnetic fields persist.  
A search for flaring activity among both active and inactive 
L dwarfs will be of interest, and should naturally occur
as 'standard' L dwarfs are reobserved.  The photometric monitoring
reported by Bailer-Jones (this volume) should show whether or
not 'starspots' exist in these dwarfs.  

Two exceptions to the general trend 
are known.  The M9.5 dwarf PC0025 was discovered due to its
strong (EW$\approx 300$\AA) H$\alpha$ emission which has
persisted for a decade\cite{pc0025discovery}.  
Is this a case of a strong 
chromosphere, or is the emission due to something else? 
It must be a rare or short-lived phenomenon, since 
no other field M dwarfs like it have been discovered.   
More recently, a 2MASS T dwarf has been discovered which has
strong, persistant H$\alpha$ emission \cite{tedwarf}.  No other
T dwarfs, or even late-L dwarfs, have such emission.  Is
this a chromosphere on a 1000K object, and if so, why 
is it unique?  Other scenarios for the emission 
are possible\cite{tedwarf}.

\section{Conclusions}

The kinematics of brown dwarfs offer a powerful way to 
measure ages.  Tangential velocities are relatively easy to obtain and should
allow the relative ages of M, L, and T dwarf samples to be 
measured.  Preliminary work has already confirmed that 
lithium L dwarfs are younger than other L dwarfs.  Age information
should help constrain models of the field brown dwarf mass
function which are sensitive to the Galactic star formation history
\cite{reidmf}.

For chromospheric activity, there is a good news and bad news.
The bad news is that activity is not a very good age indicator.
In particular, H$\alpha$ emission is more likely to be a sign
of age in field late-M and L dwarfs than youth, the opposite
of the traditional interpretation.  The good news is that 
strong trends are seen in the observations, raising interesting
questions.   Why do chromospheres --- and coronae --- disappear
at the these temperatures?  Why is there a mass dependence?  
Why does flaring persist?  What are the unusual objects
with strong, persistent emission?  The observations beg for
theoretical investigations.  As new objects are discovered and 
new observations become possible, we should expect further surprises.

With the completion of 2MASS and further follow-up, 
new tests of the scenario in this paper will be made.
2MASS has proven to be sensitive to wide L dwarf companions
of main sequence stars.  Since ages can be independently
estimated from the primary star (Kirkpatrick, this volume), 
tests of the activity-age relation will be possible.  
Additional cluster brown dwarfs will also be of great help.
2MASS is sensitive to L dwarfs in the Hyades --- while
an initial search of a portion of the cluster 
suggests that the Hyades has lost
most of its brown dwarfs\cite{grm99}, it may be hoped that at least one
L dwarf remains for 2MASS to discover.  

\section*{Acknowledgments}

I would like to thank the 2MASS Rare Objects Team (Jim Liebert,
Davy Kirkpatrick, Neill Reid, Dave Monet, Conard Dahn, 
Adam Burgasser), Mike Skrutskie, and Suzanne Hawley for 
many discussions of stars and brown dwarfs.  I acknowledge the
support of the Jet Propulsion Laboratory, California
Institute of Technology, which is operated under contract
with NASA.  This publication makes use of data products from 2MASS, 
which is a joint project of the
University of Massachusetts and IPAC/Caltech, 
funded by NASA and NSF.  I was a Visiting Astronomer, 
Kitt Peak National Observatory, National Optical Astronomy
Observatories, which is operated by the Association of Universities 
for Research in Astronomy, Inc. (AURA) under cooperative agreement 
with the NSF.

%

\end{document}